\documentclass[twocolumn]{aastex631}
\usepackage{amsmath}
\renewcommand{\fixmathlinenumbering}{}
\usepackage{hyperref}
\setcitestyle{notesep={ }}
\usepackage{bm}
\usepackage{needspace}

\graphicspath{{./}{figures/}}

\begin{document}

\title{Combining the Mass--Radius Posteriors of PSR J0030+0451 Allowing for Unknown Correlations and Model Systematics}

\correspondingauthor{Chun Huang}
\email{chun.h@wustl.edu}

\author[0009-0006-6394-2136]{Ryan O'Connor}
\affil{Physics Department and McDonnell Center for the Space Sciences, Washington University in St. Louis; MO, 63130, USA}

\author[0000-0001-6406-1003]{Chun Huang}
\affil{Physics Department and McDonnell Center for the Space Sciences, Washington University in St. Louis; MO, 63130, USA}

\author[0000-0002-4738-1168]{Alexander Y. Chen}
\affil{Physics Department and McDonnell Center for the Space Sciences, Washington University in St. Louis; MO, 63130, USA}

\begin{abstract}
The NASA Neutron star Interior Composition Explorer (\emph{NICER}) mission measures the X-ray pulse profiles of millisecond pulsars and uses pulse profile modeling (PPM) techniques to constrain their masses ($M$) and radii ($R$), in order to probe the state of matter in their interiors. One of the most studied pulsars, PSR J0030+0451, has been analyzed by multiple groups using different choices of hotspot models. The different choices of hotspot prescriptions to fit the same observational data led to different $M$--$R$ posteriors that do not completely agree with one another, resulting in a practical bottleneck for dense-matter equation-of-state (EoS) inference. In this paper, we adapt a Bayesian combination framework to the published $M$--$R$ posteriors of PSR J0030+0451 while allowing for unknown correlations and systematic uncertainties. Using this technique, we combine eight existing $M$--$R$ posteriors into a single estimate of the source parameters. The resulting constraint is $M=1.53^{+0.31}_{-0.27}\,M_\odot$, $R=13.44^{+2.95}_{-2.38}\,\mathrm{km}$, and compactness $C=0.169^{+0.018}_{-0.021}$ (68\% credible intervals). The combined distribution retains the lower- and higher-mass solutions obtained with different hotspot models. Conditional interpolation with PSR~J0437--4715 gives $R_{1.4}=11.84^{+0.88}_{-0.83}\,\mathrm{km}$. Our results provide a combined $M$--$R$ constraint for J0030+0451 that accounts for shared observations and cross-model uncertainty, and a framework for combining correlated measurements of the same source.
\end{abstract}


\section{Introduction} \label{sec:intro}
Neutron stars are probes of matter at supranuclear densities. Their core densities can reach several times the nuclear saturation density, making them natural laboratories for cold, ultra-dense matter physics \citep{Lattimer_2001,Lattimer_2007,Watts_2016,Oertel_2017,Lattimer_2021}. It is not possible to reproduce the extreme conditions in neutron star interiors in terrestrial experiments. Instead, one must infer the properties of dense matter indirectly from astronomical observables. One of the central objectives in neutron star astrophysics is therefore to measure  observables such as the stellar mass $M$, radius $R$, and dimensionless tidal deformability $\Lambda$ using multimessenger observations \citep{Lattimer_2016,Ozel_2016,Watts_2016,Abbott_2018_EOS_GW}.

For radio pulsars in bound binary systems, precise masses can often be obtained from relativistic radio-timing measurements, for example through Shapiro-delay constraints in favorable systems \citep{Stairs_2003,Demorest_2010,Cromartie_2019,Reardon_2024}. Gravitational-wave observations provide tidal-deformability constraints from GW170817 and the weaker constraint from GW190425 \citep{LIGOScientific:2017vwq,Abbott_2018_EOS_GW,GW190425}. 
For isolated neutron stars, however, obtaining precise joint constraints on mass $M$ and radius $R$ remains highly non-trivial.
The most promising approach is to model the rotationally modulated X-ray emission from their surfaces, a technique known as pulse-profile modeling (PPM). The \emph{Neutron Star Interior Composition Explorer} (\emph{NICER}) mission on the International Space Station has, for the first time, made such measurements feasible \citep{Gendreau2016,Prigozhin_16,Gendreau_2017}.  Using PPM frameworks developed and verified for \emph{NICER} data \citep{Bogdanov2019,Bogdanov2019Emission,Bogdanov2021Verification,Miller_2016,Watts_2016}, the first precise, joint $M$--$R$ constraints were obtained for PSR~J0030+0451 \citep{riley2019nicer,miller2019psr} and have since been extended to additional targets \citep{Choudhury24,Salmi24,Miller2021,Riley2021,Dittmann24}.

In practice, current PPM inference still relies on phenomenological hotspot prescriptions, because the underlying magnetospheric configuration and surface-emission physics are not yet sufficiently constrained to determine a unique predictive forward model from first principles. The stellar shape and atmosphere are included in the forward models \citep{Morsink_2007,AlGendy_14,Ho_2001,Ho_2003}. These shared ingredients are distinct from differences in hotspot geometry, background treatment, and inference settings. As a result, different hotspot prescriptions and inference pipelines can introduce model-dependent systematics and yield posterior distributions that are in measurable tension \citep{Bilous_2019,Vinciguerra24,Chen_2020,Kalapotharakos_2021,Choudhury_2024_waveform,Zhou:2025upw}.

The first \emph{NICER} target, PSR~J0030+0451 (hereafter J0030), an isolated millisecond pulsar without a precise Shapiro-delay mass measurement \citep{Bogdanov_2009}, provides a clear illustration. Two separate 2019 analyses produced broadly consistent but distinct $M$--$R$ posteriors, with \citet{miller2019psr} favoring slightly larger masses and radii than \citet{riley2019nicer}, with differences that can arise from modeling, sampling, and other pipeline choices. Subsequent studies revisited hotspot morphology and analysis systematics. \citet{Bilous_2019} interpreted the inferred hotspot arrangement as suggestive of a global multipolar magnetic field, while later reanalyses explored more flexible hotspot/temperature configurations and updated background treatments, which can shift or broaden the inferred $M$--$R$ posteriors \citep{Vinciguerra24}. More recently, \citet{Vinciguerra24} reprocessed the \emph{NICER} data and jointly modeled XMM-Newton observations to improve background constraints, demonstrating a strong sensitivity of the inferred $M$--$R$ posterior to the assumed hotspot morphology. A subsequent six-year \emph{NICER} reanalysis by \citet{Kini2026}, again performed jointly with XMM-Newton data, found that the discrepancy between viable hotspot models is reduced relative to earlier studies. These results suggest that part of the apparent tension can be mitigated by improved data quality and analysis methodology, but the source parameters remain sensitive to hotspot and other analysis assumptions \citep{Vinciguerra24,Kini2026,Choudhury_2024_waveform}. Although \emph{physics-motivated} parameterizations can reduce ad hoc flexibility, they inherit uncertainties from their underlying assumptions. The key empirical fact is that several different analyses---varying in hotspot parameterization, background treatment, and sampling implementation---provide statistically acceptable descriptions of the \emph{NICER} data, yet yield $M$--$R$ posteriors that are not fully consistent at a level relevant for dense-matter EoS inference \citep{Vinciguerra24,Kini2026}. A combined J0030 constraint therefore requires tracking these systematic contributions to the overall error budget and the dependence introduced by shared observations.

Developing improved forward models that better incorporate surface emission, magnetospheric physics, and atmospheric effects is an important long-term direction and should reduce model dependence, though it may not eliminate all systematics. In the near term, the methodological bottleneck is how to use the existing, partially discrepant posteriors without either selecting a single posterior for convenience, averaging in an ad hoc way or discarding some results as absolute outliers \citep{Lahav2000,Hobson2002}. Our goal therefore is to combine the currently available J0030 $M$--$R$ posteriors into a single estimate of its mass and radius, allowing for unknown correlations and systematics so that the final constraint retains the cross-analysis disagreement.

To this end, we adopt the Bayesian good/bad mixture framework proposed by \citet{press1996} for combining potentially discrepant measurements, with measurement-specific shift and broadening parameters in the bad component inspired by \citet{Bernal_2018}. Related approaches using continuous hyperparameters for dataset weighting were developed by \citet{Lahav2000} and \citet{Hobson2002}. The Press framework has also been applied in stellar astrophysics by \citet{phillips2025}. In this framework, each input posterior distribution is assigned an additional probability of being ``good'' (its error bar is reliable) or ``bad'' (its error bar underestimates the systematic uncertainties). A joint distribution of the underlying physical parameters is constructed from the input posteriors, with additional shifts and broader uncertainties in the bad components. Note that this ``good/bad'' designation is not a statement about the physical correctness of the given model, but simply a statistical statement of whether the data suggests that there are systematic uncertainties that are not accounted for in the original model.

We apply this methodology to existing \emph{NICER}-informed $M$--$R$ posteriors for PSR~J0030+0451. In the original \citet{press1996} formulation, both the good and bad components are modeled as Gaussians. However, published PPM posteriors are often noticeably non-Gaussian, exhibiting skewness, extended tails, or multi-modality. We therefore replace the good component for each measurement with a non-parametric kernel density estimate (KDE) constructed directly from the published posterior samples, while retaining a broadened and shifted Gaussian for the bad component. This preserves the non-Gaussian structure of each published posterior. Because the analyses share observations, we temper their combined contribution to allow for unknown dependence and include a common systematic displacement. We carry out the combination in mass and compactness and derive radius from the joint distribution. The resulting posterior summarizes the J0030 $M$--$R$ values allowed by the selected analyses and quantifies the agreement of each individual input with the combined constraint.

The paper is organized as follows. In Section~\ref{sec:method} we describe in detail the statistical framework used to combine the existing J0030 inference results. Section~\ref{sec:result} presents the resulting compactness and mass--radius posteriors, together with the correlation comparison and per-measurement consistency scores. Section~\ref{sec:conclusion} summarizes our conclusions and discusses the comparison with other neutron stars and the derived canonical radius.

\section{Methods}\label{sec:method}

\Needspace{6\baselineskip}
\subsection{Published posteriors and KDE representation}\label{sec:method-extension}
Let $q_i(\boldsymbol{x})$ be the normalized density reconstructed from the posterior samples of analysis $i$, with $i=1,\ldots,N$ and $N=8$. We use $\boldsymbol{x}=(M,C)^T$, with $C=GM/(Rc^2)$. We transform each published mass--radius sample to $(M,C)$ before constructing its KDE. All mixtures, priors, shifts, and systematic covariances are defined in this space. Radius is derived after pooling. All radii are equatorial radii. The input objects are posteriors, not likelihoods from independent experiments. If the original prior is $\pi_i(\boldsymbol{x})$, then $q_i(\boldsymbol{x})\propto L_i(\boldsymbol{x})\pi_i(\boldsymbol{x})$. A likelihood calculation would require division by that prior over its support. We pool the published posterior densities, retaining the original analyses' prior dependence, and apply one specified source prior $\pi_x$ to the pooled density. This is a generalized pooling prescription.

For posterior samples $\{\boldsymbol{x}_{ij}\}_{j=1}^{n_i}$, the Gaussian KDE is
\begin{equation}
q_i(\boldsymbol{x})=\frac{1}{n_i}\sum_{j=1}^{n_i}
\mathcal{N}_2(\boldsymbol{x};\boldsymbol{x}_{ij},\boldsymbol{H}_i),
\label{eq:KDE}
\end{equation}
where $\boldsymbol{H}_i=h_i^2\widehat{\boldsymbol{V}}_i$ and $\widehat{\boldsymbol{V}}_i$ is the sample covariance matrix.

We implement the KDE using {the Python module} \texttt{scipy.stats.gaussian\_kde}, and choose the bandwidth according to Silverman’s suggestion for multivariate data \citep{silverman1986}:
\begin{equation}
    h_i \;=\; \left(\frac{n_i(d+2)}{4}\right)^{-\frac{1}{d+4}},
\end{equation}
where $n_i$ is the number of posterior samples for input $i$ and $d=2$ for the joint $(M,C)$ density.  We set \texttt{bw\_method='silverman'} explicitly. The KDE bandwidth controls numerical smoothing and is separate from the systematic broadening below.

\Needspace{6\baselineskip}
\subsection{Measurement-specific shifts and broadening}\label{sec:method-original}
Each analysis contributes a mixture of its published KDE and a shifted, broadened Gaussian. Before marginalization, the bad component of input $i$ is $\mathcal{N}_2(\boldsymbol{x};\boldsymbol{\mu}_i+\boldsymbol{\Delta}_i,\alpha_i^2\boldsymbol{\Sigma}_{B,i})$, where $\boldsymbol{\mu}_i=(\mu_{M,i},\mu_{C,i})^T$ contains the marginal medians, $\alpha_i\geq1$ is a width factor, and $\boldsymbol{\Delta}_i=(\Delta_{M,i},\Delta_{C,i})^T$ is a displacement. The displacement has a zero-mean Gaussian prior, $\boldsymbol{\Delta}_i\sim\mathcal{N}_2(0,\boldsymbol{\Sigma}_{\sigma})$ with $\boldsymbol{\Sigma}_{\sigma}=\mathrm{diag}(\sigma_M^2,\sigma_C^2)$. Integrating over $\boldsymbol{\Delta}_i$ analytically and over $\alpha_i$ numerically gives the mixture density of input $i$,
\begin{align}
\ell_i(\boldsymbol{x}\mid p,\boldsymbol{\sigma})
&=p\, q_i(\boldsymbol{x})+(1-p)\bar B_i(\boldsymbol{x}\mid\boldsymbol{\sigma}),\label{eq:mixture}\\
\bar B_i(\boldsymbol{x}\mid\boldsymbol{\sigma})&=\int_1^{\infty}d\alpha_i\,\pi(\alpha_i)\,
\mathcal{N}_2(\boldsymbol{x};\boldsymbol{\mu}_i,
\alpha_i^2\boldsymbol{\Sigma}_{B,i}+\boldsymbol{\Sigma}_{\sigma}),\label{eq:goodandbad}
\end{align}
where $\boldsymbol{\sigma}=(\sigma_M,\sigma_C)$. As in \citet{press1996}, a single mixing probability $p$ is shared by all inputs: it is the prior probability that any one input belongs to the good component. The shift scales $\boldsymbol{\sigma}$ are likewise common to all inputs and are inferred hierarchically, while each input has its own $\alpha_i$ and $\boldsymbol{\Delta}_i$. The shared nuisance parameters are therefore $\boldsymbol{\eta}=(p,\sigma_M,\sigma_C)$. The terms ``good'' and ``bad'' refer to these two statistical components. They do not classify hotspot morphologies as physically correct or incorrect. The good component retains the published uncertainty; the bad component permits a displacement and additional width.

The marginal baseline widths have a simple quantile definition. If $Q_{i,a}(u)$ is the quantile of coordinate $a\in\{M,C\}$, define
\begin{align}
w_{i,a}&=\tfrac12[Q_{i,a}(0.995)-Q_{i,a}(0.005)],\nonumber\\
S_{i,a}&=\frac{w_{i,a}}{\Phi^{-1}(0.84)},\label{eq:S_calibration}
\end{align}
where $\Phi$ is the standard normal cumulative distribution. Thus the central 68\% width of a baseline bad Gaussian matches the central 99\% width of the corresponding input marginal. This matches interval widths; for a skewed posterior it need not match both endpoints about the median.

The bad component uses a covariance matrix $\boldsymbol{\Sigma}_{B,i}$ which is diagonal in $(M,C)$ and has widths following Eq.~\eqref{eq:S_calibration}.  Because the combination is done in $(M,C)$, this diagonal bad component is still correlated in $(M,R)$: a shift or broadening in mass at approximately fixed compactness gives $R=GM/(Cc^2)\propto M$, so the bad components are elongated along lines of constant compactness in $(M,R)$ space. There exists correlation between $M$ and $C$ in the input posteriors, which is modeled by their ``good'' distributions $q_i$, but those off-diagonal components are not modeled in the broadened, simplified ``bad'' distribution.

The priors are $p\sim\mathrm{Uniform}(0,1)$ and $\pi(\alpha_i)\propto\exp[-(\alpha_i-1)]$, i.e.\ $\alpha_i-1\sim\mathrm{Exponential}(1)$, truncated at $\alpha_i=10$ where the omitted prior mass is $e^{-9}$. Each shift scale has a log-uniform prior, $\pi(\sigma_a)\propto1/\sigma_a$, with $10^{-4}\leq\sigma_C\leq10^{-0.5}$ and $10^{-4}\leq\sigma_M/M_\odot\leq1$. Because this prior is normalizable only between such bounds, the bounds are part of the prior specification. We evaluate $p$ on 25 equally spaced points, $\alpha_i$ on 40 points, and each $\sigma_a$ on 6 logarithmically spaced points, and marginalize by trapezoidal quadrature. The scales determine how far an individual posterior may move and how much it may broaden; they are assumptions about unresolved systematics. Restricting $\alpha_i\geq1$ ensures that the width factor broadens its baseline Gaussian. It does not impose a Gaussian approximation on $q_i$.

\Needspace{6\baselineskip}
\subsection{Correlation tempering}\label{sec:tempering}

The analyses share observations, so their mixture densities cannot be multiplied as independent measurements. We define a tempered pool at fixed shared nuisance parameters $\boldsymbol{\eta}=(p,\boldsymbol{\sigma})$,
\begin{align}
L_\rho(\boldsymbol{x}\mid\boldsymbol{\eta})
&=\prod_{i=1}^{N}\ell_i(\boldsymbol{x}\mid p,\boldsymbol{\sigma})^{\kappa(\rho)w_i},\label{eq:tempered_pool}\\
\kappa(\rho)&=\frac{1}{1+(N-1)\rho},\quad 0\leq\rho\leq1,\label{eq:kappa}
\end{align}
with $w_i\geq0$ and $\sum_iw_i=N$. Thus $\log L_\rho=\kappa\sum_iw_i\log\ell_i$; there is no additional factor of $N$ under this weight convention. Equal weights mean $w_i=1$. The effective information count is $N_{\rm eff}=N\kappa$. At $\rho=0$ the independent-product endpoint is recovered. At $\rho=1$ the total exponent is one, giving a geometric pool with the information scale of one measurement. This endpoint still depends on the selected input distributions. The input-specific parameters $\alpha_i$ and $\boldsymbol{\Delta}_i$ are integrated inside $\bar B_i$ (Eq.~\ref{eq:goodandbad}) before tempering, so the exponent acts on each input's marginal mixture density. For $\kappa<1$ this differs from tempering at fixed $(\alpha_i,\boldsymbol{\Delta}_i)$ and integrating afterwards, because a power of an average is not the average of the powers. The shared parameters $\boldsymbol{\eta}$ are integrated after the product, as in Eq.~\eqref{eq:final_pool}.

For identical mixture densities $\ell_i=\ell$, the pool is $L_\rho=\ell^{N_{\rm eff}}$. The $\rho=1$ endpoint returns that density, while the $\rho=0$ endpoint raises it to the $N$th power. This limiting case shows how the prescription removes the repeated-information contraction at full correlation. The factor also has an exact interpretation for equal-variance Gaussian measurements with a common pairwise correlation. If their covariance matrix has diagonal entries $\sigma^2$ and off-diagonal entries $\rho\sigma^2$, the variance of their mean is $\sigma^2[1+(N-1)\rho]/N$. Applying Eq.~\eqref{eq:kappa} to the independent Gaussian product gives this variance. Published PPM posteriors have unequal widths and non-Gaussian shapes, so we use this relation as an effective information prescription. A scalar $\rho$ does not reconstruct their full cross-analysis covariance.

For the reported mass--radius summaries we prescribe $\pi_\rho=\mathrm{Uniform}(0,1)$. This admits information counts from one to eight. The prior is a sensitivity choice, not an estimate of photon overlap. It has density $N/[ (N-1)N_{\rm eff}^2]$ in $N_{\rm eff}$ and therefore is not uniform in the effective number of measurements. Priors such as $\mathrm{Beta}(4,1)$ and $\mathrm{Beta}(1,4)$ emphasize higher and lower effective correlations, respectively; they define additional sensitivity calculations, not results reported here.

\paragraph{Density-agreement diagnostic.}
We form a separate effective-correlation diagnostic by normalizing each powered mass--compactness mixture before evaluating their overlap. At fixed nuisance parameters define
\begin{align}
t_{i,\rho}(\boldsymbol{x})&=
\frac{\ell_i(\boldsymbol{x})^{\kappa w_i}}
{\int d^2 u\,\ell_i(\boldsymbol{u})^{\kappa w_i}},\nonumber\\
Z_{\rm ov}(\rho,\boldsymbol{\eta})&=\int d^2 x\,\prod_i t_{i,\rho}(\boldsymbol{x}),\nonumber\\
p_{\rm diag}(\rho\mid D)&\propto\pi_\rho(\rho)\int d\boldsymbol{\eta}\,\pi_\eta(\boldsymbol{\eta})Z_{\rm ov}(\rho,\boldsymbol{\eta}).
\label{eq:rho_diag}
\end{align}
This normalization makes the relative diagnostic weights independent of density units. The common systematic covariance does not enter this calculation. The diagnostic measures agreement under the powered-density model. Eight identical analyses of the same photons can agree just as closely as eight independent measurements, so density agreement does not determine the physical correlation of the observations. We use this diagnostic separately from the prescribed correlation weights in Eq.~\eqref{eq:final_pool}.

\Needspace{6\baselineskip}
\subsection{Common systematic displacement and the final pool}\label{sec:groupsys}
Measurement-specific shifts allow the published solutions to disagree. They do not account for an error shared by all analyses. We represent a common displacement by $\boldsymbol{b}\sim\mathcal{N}_2(0,\boldsymbol{\Sigma}_{\rm sys})$ and convolve the tempered pool:
\begin{equation}
\widetilde L_\rho(\boldsymbol{x}\mid\boldsymbol{\eta},\boldsymbol{\Sigma}_{\rm sys})
=\int d^2 x'\,\mathcal{N}_2(\boldsymbol{x}-\boldsymbol{x}';0,\boldsymbol{\Sigma}_{\rm sys})
L_\rho(\boldsymbol{x}'\mid\boldsymbol{\eta}).
\label{eq:systematic_convolution}
\end{equation}
For $(M,C)$ we write
\begin{equation}
\boldsymbol{\Sigma}_{\rm sys}=
\begin{pmatrix}
\sigma_{M,\rm sys}^2&0\\
0&\sigma_{C,\rm sys}^2
\end{pmatrix},\label{eq:syscov}
\end{equation}
with independent $\sigma_{a,\rm sys}\sim\mathrm{HalfNormal}(s_a)$ and $s_a=\mathrm{Std}_i(\mu_{i,a})$, each evaluated on 15 equally spaced points up to $4s_a$. Taking $\boldsymbol{\Sigma}_{\rm sys}$ diagonal in $(M,C)$ lets the two-dimensional convolution factor into one-dimensional convolutions. A common mass displacement at fixed compactness still maps to a correlated mass--radius displacement, $\delta R/R=\delta M/M$. For the supplied ensemble, $s_M=0.225\,M_\odot$ and $s_C=0.0118$. These scatter-based scales connect the allowed common displacement to the spread of the supplied analyses. They do not measure a common bias: all analyses could share a bias that their scatter does not reveal.

To specify the averaging convention, first marginalize the nuisance parameters at fixed $\rho$:
\begin{align}
A_\rho(\boldsymbol{x})&=\pi_x(\boldsymbol{x})\int d\boldsymbol{\eta}\,d\boldsymbol{\Sigma}_{\rm sys}\,
\pi_\eta\pi_{\rm sys}\widetilde L_\rho(\boldsymbol{x}),\nonumber\\
Z_\rho&=\int d^2 x\,A_\rho(\boldsymbol{x}),\nonumber\\
P_{\rm pool}(\boldsymbol{x}\mid D)&=\int_0^1d\rho\,\pi_\rho(\rho)
\frac{A_\rho(\boldsymbol{x})}{Z_\rho}.\label{eq:final_pool}
\end{align}
Here $D$ denotes the collection of published posterior samples. The source prior $\pi_x$ is uniform in $(M,C)$ over the computational domain described below; in mass--radius coordinates this corresponds to a density proportional to $M/R^2$. Equation~\eqref{eq:final_pool} averages normalized conditional distributions with the prescribed correlation weights. It does not update those weights with the diagnostic in Section~\ref{sec:tempering}. Normalization at fixed $\rho$ also removes the arbitrary density-unit factor from raising a density to a variable power. Numerical peak rescaling within the nuisance integral must retain the removed scale factors; otherwise the nuisance weights change. Because the convolution is linear and $\pi_{\rm sys}$ does not depend on $\rho$ or $\boldsymbol{\eta}$, we apply it once to the $\rho$-averaged pool. The domain has no margin beyond the input samples, so for each value of $\boldsymbol{\Sigma}_{\rm sys}$ we renormalize the convolved density on the domain: probability broadened past the domain edges is discarded rather than retained.

For presentation, samples pooled in $(M,C)$ are mapped to $(M,R)$ using
\begin{equation}
R\,[\mathrm{km}]=1.476625\frac{M/M_\odot}{C}.
\label{eq:radius_transform}
\end{equation}
The mass--radius density is obtained from the mass--compactness density using the Jacobian $P_{MR}(M,R)=P_{MC}(M,C)C/R$. Transforming the pooled samples includes this Jacobian automatically. This transformation presents the result in mass--radius coordinates; the pooling and systematic model remain defined in $(M,C)$.

The archived mass--compactness densities are represented on an $80\times80$ grid with $0.95\leq M/M_\odot\leq2.20$ and $0.10036\leq C\leq0.21220$; the correlation grid contains 21 equally spaced values from zero to one. We normalize the stored density on this domain and use bilinear interpolation within its cells. Marginal percentiles are evaluated by integrating this density, including the Jacobian when deriving radius. Sampling the same interpolant preserves the joint distribution for the conditional propagation in Section~\ref{sec:conclusion}. These summaries are conditional on the archived numerical domain; no probability is assigned outside it.

\Needspace{6\baselineskip}
\subsection{Model weights and evidence}\label{sec:weights}
Equal weights define the analysis-choice sensitivity calculation reported here. They do not assert equal fit quality or equal posterior probabilities for the hotspot models. In particular, the joint-data evidence comparison of \citet{Vinciguerra24} disfavors ST--U and ST+PST relative to the more flexible temperature models. Keeping those distributions in the ensemble tests how much the inferred source parameters depend on adopting the published alternatives. It gives the result a different interpretation from evidence-weighted model averaging.

For models $i$ fitted to the same data within an analysis group $g$, evidence weights would be
\begin{align}
a_{i\mid g}&=\frac{\pi(i\mid g)\mathcal{Z}_{i,g}}
{\sum_{j\in g}\pi(j\mid g)\mathcal{Z}_{j,g}},\nonumber\\
\mathcal{Z}_{i,g}&=\int L_{i,g}(\boldsymbol{\vartheta})\pi_{i,g}(\boldsymbol{\vartheta})d\boldsymbol{\vartheta}.
\label{eq:evidence_weights}
\end{align}
Maximum log likelihood is not Bayesian evidence. With group weights $W_g$ satisfying $\sum_gW_g=1$, the pooling exponents can be set by $w_i=NW_g a_{i\mid g}$. Evidence values from different data selections cannot be globally normalized into model probabilities. Nor does weighting within a group remove photon overlap between groups. An arithmetic model average $\sum_i a_{i\mid g}q_i$ answers which source parameters are supported after averaging model probabilities within that group; the tempered product answers a consensus question under an assumed effective information count. We report the equal-weight construction explicitly and do not assign evidence weights to the reported results.

\Needspace{6\baselineskip}
\subsection{Input analyses}\label{sec:dataset}
The eight mass--radius posteriors used in our analysis are drawn from four published NICER studies of PSR~J0030+0451: \citet{riley2019nicer}, \citet{Vinciguerra24}, \citet{miller2019psr}, and \citet{Kini2026}. The currently available posteriors for J0030 differ greatly in their hotspot parameterizations, inference configurations, and background treatments, among other modeling choices. They are eight analyses of overlapping NICER and XMM-Newton observations, not eight independent data sets.

In \citet{riley2019nicer}, the ST+PST configuration is presented as the headline model. In this scenario, the surface of PSR~J0030+0451 is modeled with two hotspots located on the same rotational hemisphere (the hemisphere tilted away from the observer). One hotspot is a relatively small, nearly circular region subtending only a few degrees, while the other is an extended, narrow crescent region. Although ST+PST is adopted as the preferred configuration in that work, \citet{riley2019nicer} emphasize that the available data do not provide decisive evidence to distinguish among several alternative hotspot morphologies they considered. In this paper, we use the published ST+PST posterior from \citet{riley2019nicer} as one of our eight input models.

\citet{Vinciguerra24} perform an updated analysis of the same NICER data, jointly modeled with XMM-Newton observations to improve background characterization. Their study revisits and extends the hotspot morphologies of \citet{riley2019nicer}, allowing for more hotspot configurations. From \citet{Vinciguerra24}, we include four posterior distributions: ST+PST, ST+PDT, PDT--U, and ST--U. The ST--U model describes a hotspot configuration with two circular emitting regions of uniform temperature with independent parameters. In contrast, the ST+PST and ST+PDT models feature a circular primary hotspot plus an additional, more complex secondary region. The PDT--U model further generalizes this picture by allowing both hotspots to consist of two circular emitting zones with different temperatures, effectively introducing a temperature distribution across each hotspot. \citet{Vinciguerra24} find that the PDT--U configuration is significantly preferred by the Bayesian evidence. Their joint-data evidence disfavors ST--U and ST+PST relative to the more flexible temperature models. Equal weights in our pool define an analysis-choice sensitivity calculation, as specified in Section~\ref{sec:weights}.

\citet{miller2019psr} present two additional models for the surface emission of PSR~J0030+0451. Their preferred configuration consists of three oval-shaped, uniform-temperature hotspots (which we denote as the ``3spot'' model). They also report a ``2spot'' model with two uniform-temperature hotspots that, in terms of both waveform reproduction and inferred $(M,R)$, is statistically indistinguishable from the 3spot configuration. This near-degeneracy is reflected in the close overlap of their $M$--$R$ posterior contours. In our analysis, we include both the 2spot and 3spot posteriors from \citet{miller2019psr} as two further analyses of the same NICER observations.

We also include one posterior from the recent six-year NICER reanalysis of PSR~J0030+0451 by \citet{Kini2026}. That study incorporates NICER observations obtained between 2017 July and 2023 January, increasing the number of X-ray counts by about 50\% relative to previous analyses, and jointly models them with the same archival XMM-Newton observations used in \citet{Vinciguerra24}. In that work, the phase- and energy-resolved emission is modeled using the two hotspot geometries identified by \citet{Vinciguerra24} as viable descriptions of the data, namely ST+PDT and PDT--U. They identified PDT--U as a better model supported by the Bayesian evidence. We therefore include the updated PDT--U posteriors from \citet{Kini2026} as one additional input in our analysis. 

In summary, the eight models considered here comprise: one ST+PST posterior from \citet{riley2019nicer}; four posteriors (ST+PST, ST+PDT, PDT--U, and ST--U) from \citet{Vinciguerra24}; two posteriors (2spot and 3spot) from \citet{miller2019psr}; and one updated posterior (PDT--U) from \citet{Kini2026}.  Together, they span a representative range of physically motivated hotspot morphologies and analysis choices that are all capable of reproducing the NICER (and, where applicable, XMM-Newton) data for PSR~J0030+0451. Our Bayesian combination is constructed to be agnostic about which morphology is \emph{correct}, and instead summarizes the mass and radius allowed by the ensemble of analyses.

\section{Results}\label{sec:result}
Published NICER pulse-profile inferences for PSR~J0030+0451 exhibit model-dependent scatter in the $M$--$R$ plane that is large enough to affect downstream EoS conclusions. Different hotspot prescriptions yield overlapping yet systematically shifted credible regions. We combine these published posteriors into a single J0030 constraint while accounting for their shared observations and model systematics.

In this section, we present the combined compactness and mass--radius results for PSR~J0030+0451. Alongside the combined posteriors, we report a consistency score for each individual measurement. This ranking is a quantitative indicator of how statistically consistent each measurement is with the ensemble within our framework. All quoted intervals are the 16th and 84th marginal percentiles about the median. The plotted two-dimensional contours are highest-density regions (HDRs); their projections need not coincide with the marginal intervals.

\Needspace{6\baselineskip}
\subsection{Compactness}\label{sec:compactness_results}
\begin{figure}[!t]
\centering
\includegraphics[width=\columnwidth]{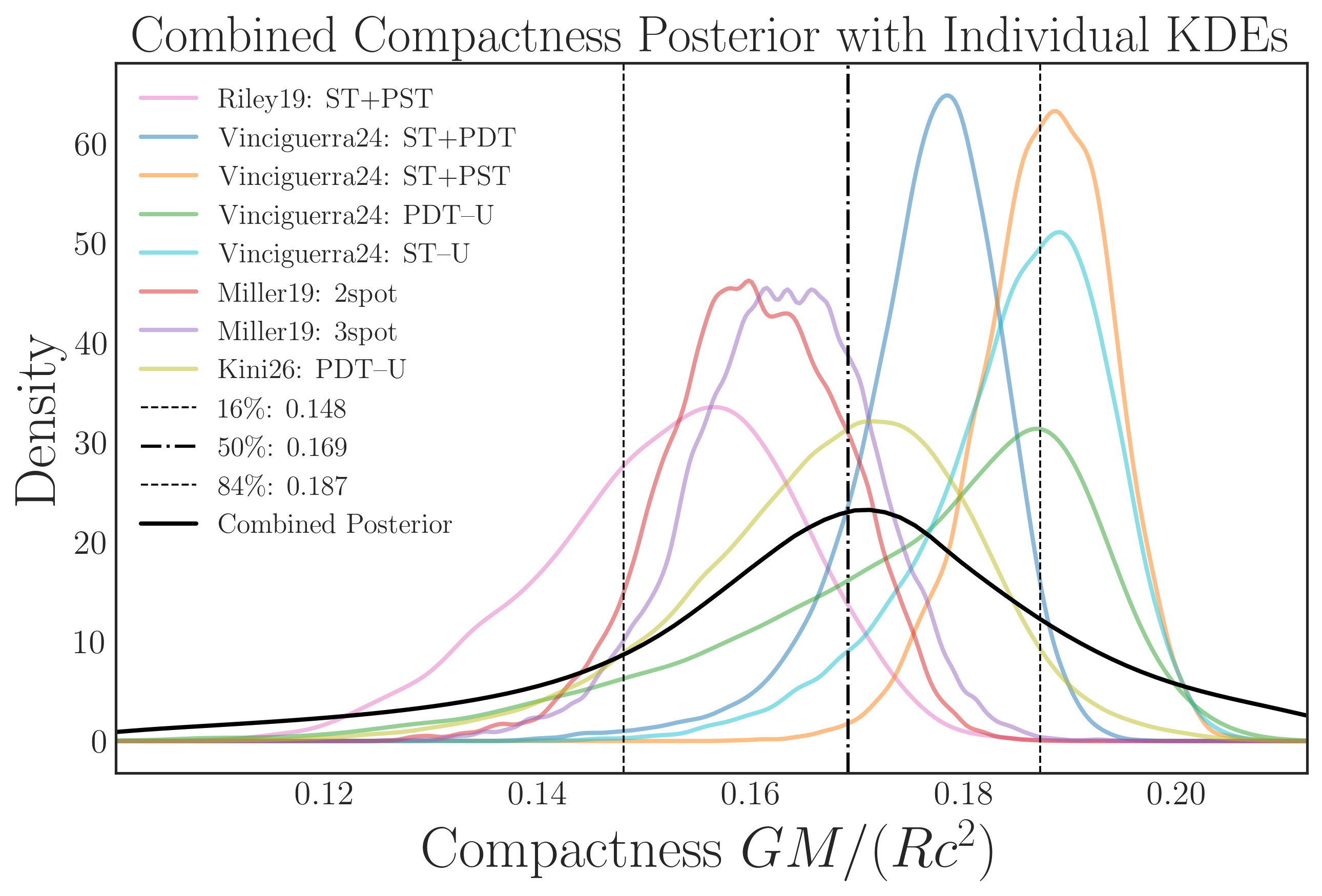}
\caption{Compactness marginal of the adopted joint mass--compactness pool (black), compared with the KDEs of the eight input analyses (colored). Vertical lines mark the 16th, 50th, and 84th percentiles, approximately 0.148, 0.169, and 0.187. The pooled interval retains the spread among the lower- and higher-compactness input solutions.}
\label{fig:KDECompactness}
\end{figure}

In Figure~\ref{fig:KDECompactness}, we display the posterior distributions of the compactness, derived from the posterior samples of each model. These distributions are shown with fainter colors in the background, computed using the KDE method. As discussed in Section~\ref{sec:dataset}, we consider eight distinct posteriors in total. Their compactness peaks span approximately 0.15--0.19, substantially broader than the typical width of an individual measurement. We use Silverman's rule of thumb for the input KDE bandwidths, as described in Section~\ref{sec:method-extension}.

The set of models includes one posterior from \citet{riley2019nicer} (denoted as Riley19 in the plot), corresponding to their ST+PST headline result; four posteriors from \citet{Vinciguerra24} (denoted as Vinciguerra24); two posteriors from \citet{miller2019psr} (denoted as Miller19); and one updated posterior from \citet{Kini2026} (denoted as Kini26). Further details on each model and the rationale for selecting these representative data sets are provided in Section~\ref{sec:dataset}.

We present the combined compactness posterior for PSR~J0030+0451 in Figure~\ref{fig:KDECompactness} with the solid black curve. The combined result is
\begin{equation}
C=0.169^{+0.018}_{-0.021},
\label{eq:compactness_result}
\end{equation}
with central interval $[0.148,0.187]$. A direct comparison between this combined compactness and the individual measurements used as inputs is summarized in Table~\ref{tab:summary_table}. Its median is close to the updated PDT--U result from \citet{Kini2026}, while its broad interval retains both the lower-compactness Riley19 and Miller19 solutions and the higher-compactness Vinciguerra24 solutions.

The compactness result is obtained by marginalizing the joint mass--compactness posterior over mass. The mass and radius results below are derived from this same joint distribution, retaining their dependence on compactness. In particular, the compactness median is not the ratio of the marginal mass and radius medians.

\begin{table*}[!t]
\centering
\begin{tabular}{lccc}
\hline\hline
J0030 input analysis & Compactness & $M/M_\odot$ & $R$ (km)\\
\hline
\citet{riley2019nicer}: ST+PST & $0.153^{+0.010}_{-0.014}$ & $1.29^{+0.17}_{-0.16}$ & $12.57^{+1.37}_{-1.29}$\\
\citet{Vinciguerra24}: ST+PST & $0.188^{+0.006}_{-0.009}$ & $1.93^{+0.10}_{-0.12}$ & $15.23^{+0.52}_{-0.84}$\\
\citet{Vinciguerra24}: ST+PDT & $0.177^{+0.006}_{-0.007}$ & $1.40^{+0.13}_{-0.12}$ & $11.71^{+0.87}_{-0.83}$\\
\citet{Vinciguerra24}: PDT--U & $0.179^{+0.011}_{-0.022}$ & $1.70^{+0.18}_{-0.19}$ & $14.43^{+0.88}_{-1.05}$\\
\citet{Vinciguerra24}: ST--U & $0.186^{+0.007}_{-0.010}$ & $1.88^{+0.13}_{-0.19}$ & $15.12^{+0.64}_{-1.31}$\\
\citet{miller2019psr}: 2 spots & $0.160^{+0.009}_{-0.008}$ & $1.44^{+0.18}_{-0.16}$ & $13.27^{+1.30}_{-1.15}$\\
\citet{miller2019psr}: 3 spots & $0.163^{+0.008}_{-0.009}$ & $1.44^{+0.15}_{-0.15}$ & $13.01^{+1.25}_{-1.06}$\\
\citet{Kini2026}: PDT--U & $0.169^{+0.011}_{-0.014}$ & $1.43^{+0.20}_{-0.17}$ & $12.68^{+1.31}_{-1.04}$\\
This work (adopted) & $0.169^{+0.018}_{-0.021}$ & $1.53^{+0.31}_{-0.27}$ & $13.44^{+2.95}_{-2.38}$\\
\hline
\end{tabular}
\caption{Mass, radius, and compactness summaries for the eight supplied input distributions and the adopted pool. Uncertainties are central 68\% marginal intervals. The adopted row is evaluated from the continuous bilinear mass--compactness density, with radius derived from the same joint distribution.}
\label{tab:summary_table}
\end{table*}

\Needspace{6\baselineskip}
\subsection{Mass--radius summary}\label{sec:2D_results}
\begin{figure*}[!t]
\centering
\includegraphics[width=\textwidth]{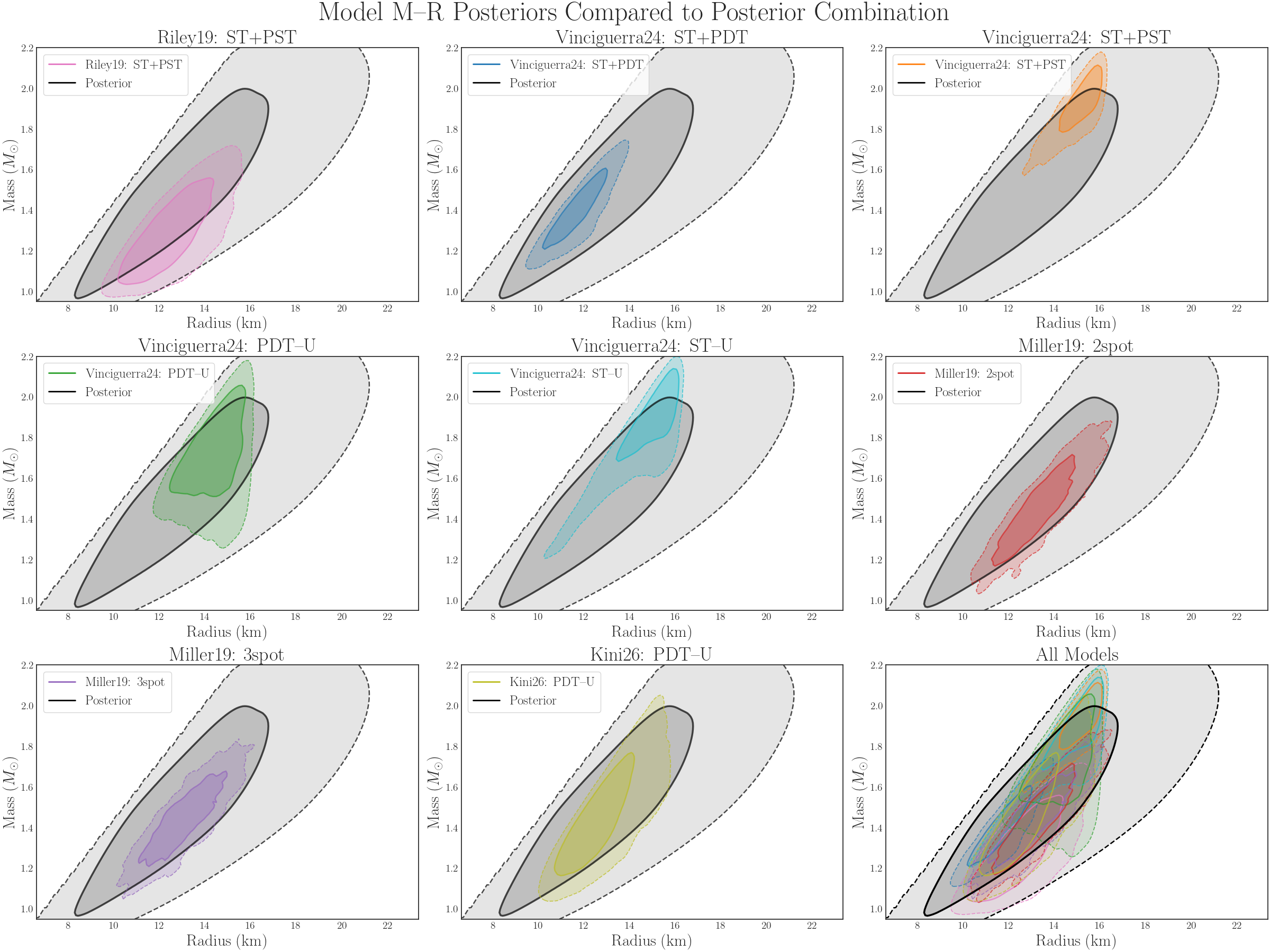}
\caption{Individual mass--radius posteriors for J0030 compared with the adopted mass--compactness pool transformed to $(M,R)$. Colored contours show each input; gray regions and black contours show the pooled summary. Solid and dashed contours indicate 68\% and 95\% HDRs. The final panel overlays all inputs. The broad pool overlaps the lower- and higher-mass solution groups and extends beyond their individual contours. Parts of the pooled tails lie outside the displayed axes.}
\label{fig:3x3PosteriorvsIndividual}
\end{figure*}

In Figure~\ref{fig:3x3PosteriorvsIndividual} we show the overall spread of the mass--radius posteriors obtained from the different hotspot models in various colors. The input masses extend from approximately $1\,M_\odot$ to $2.2\,M_\odot$, while the principal radius solutions lie near 10--16~km. Such dispersion reflects differences in background modeling, hotspot morphology, sampling, and other analysis choices.

\begin{figure}[!t]
\centering
\includegraphics[width=\columnwidth]{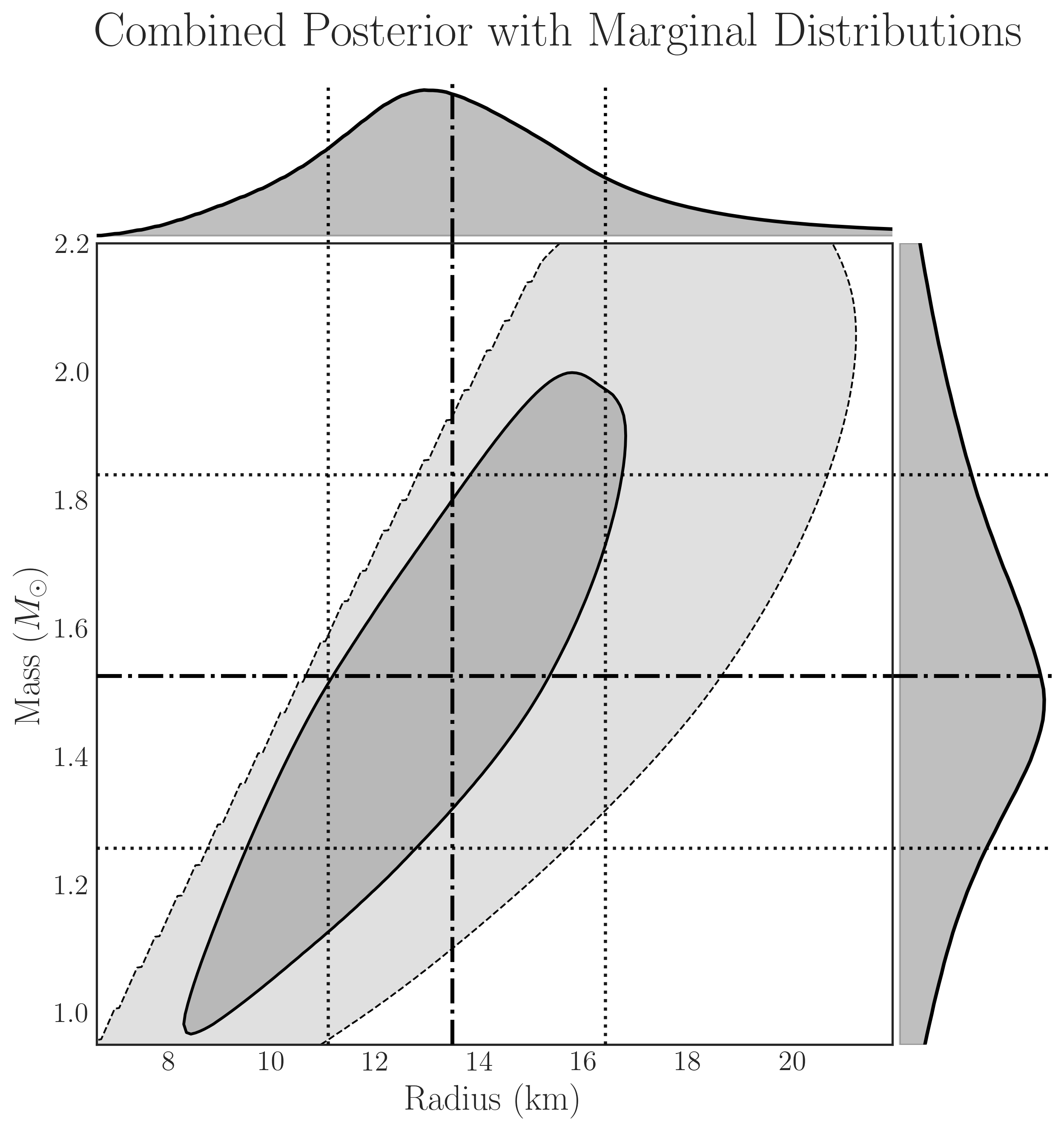}
\caption{Adopted mass--radius summary derived from the joint mass--compactness pool. The solid contour and darker shading indicate the 68\% HDR; the dashed contour and lighter shading indicate the 95\% HDR. The upper and right panels show the radius and mass marginals. Dash-dotted guides mark marginal medians and dotted guides mark the 16th and 84th percentiles. The long radius tail and positive mass--radius correlation are retained in the joint distribution.}
\label{fig:MarginalPosteriors}
\end{figure}

We overlay our combined posterior (black) on the individual mass--radius posteriors; the combined posterior and its one-dimensional marginal distributions are shown in Figure~\ref{fig:MarginalPosteriors}. From the combination with a uniform correlation prior and common systematics, we infer
\begin{align}
M&=1.53^{+0.31}_{-0.27}\,M_\odot,\nonumber\\
R&=13.44^{+2.95}_{-2.38}\,\mathrm{km}.
\label{eq:adopted_result}
\end{align}
The central intervals are $[1.26,1.84]\,M_\odot$ and $[11.06,16.38]$~km. Compared to the individual NICER-based measurements, the combined result points to a mass slightly above the canonical $1.4\,M_\odot$, while remaining consistent with $1.4\,M_\odot$ within the central 68\% interval.

Figure~\ref{fig:3x3PosteriorvsIndividual} compares the transformed pool with every input. The gray region follows the positive mass--radius direction occupied by the published solutions. It overlaps both the lower-mass Riley19 and Miller19 solutions and the higher-mass, higher-radius Vinciguerra24 solutions. At approximately fixed compactness, a broader mass distribution extends the allowed radius through $R\propto M$. This geometry explains why the pool is elongated and why retaining the high-mass alternatives also retains large radii. The result summarizes the spread of source solutions allowed by the selected analyses.

The joint distribution and its marginals are shown in Figure~\ref{fig:MarginalPosteriors}. The radius marginal is skewed toward large radii. Its central 95\% interval is $[8.79,21.47]$~km, while the corresponding mass interval is $[1.03,2.10]\,M_\odot$. These are marginal intervals, not rectangular boundaries of a joint 95\% HDR. The marginal median guides cross inside the elongated source region, but the mass and radius are strongly dependent. A pair of one-dimensional intervals cannot represent that dependence.
\begin{figure*}[!t]
\centering
\includegraphics[width=\textwidth]{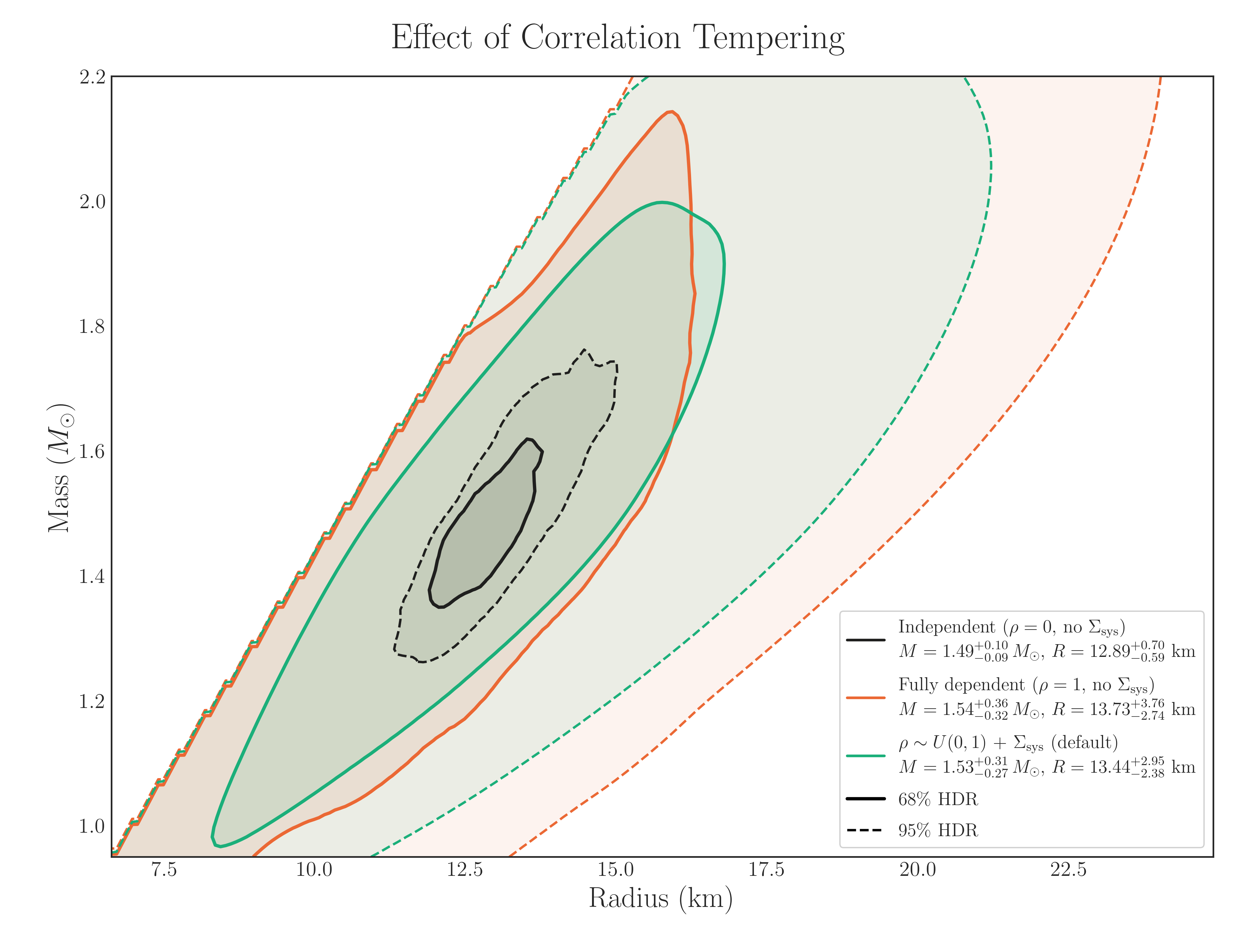}
\caption{Effect of correlation tempering on the joint mass--compactness pool, displayed in mass--radius coordinates. Black curves show $\rho=0$ and orange curves show $\rho=1$, both without a common systematic displacement. Green curves show the adopted uniform-correlation average with common systematics marginalized. Solid and dashed contours indicate 68\% and 95\% HDRs, and the legend lists 68\% credible region of the marginal constraints in $M$ and $R$. All three calculations pool in $(M,C)$.}
\label{fig:MC_tempering}
\end{figure*}

\Needspace{6\baselineskip}
\subsection{Effect of correlation tempering}\label{sec:temper_results}
To assess how unknown dependence affects this J0030 estimate, Figure~\ref{fig:MC_tempering} compares three information prescriptions in the same mass--compactness construction. The black curves are the independent-product limit, $\rho=0$, with no common systematic displacement. They give $M=1.49^{+0.10}_{-0.09}\,M_\odot$ and $R=12.89^{+0.70}_{-0.59}$~km. The orange curves set $\rho=1$ while keeping the common systematic covariance zero. This reduces the ensemble's total exponent from eight to one and gives $M=1.54^{+0.36}_{-0.32}\,M_\odot$ and $R=13.73^{+3.76}_{-2.74}$~km. The endpoint comparison isolates the effect of tempering: the central radius width increases from 1.30 to 6.50~km, a factor of about 5.0.

The green curves show the adopted uniform-correlation average with common systematics. Its central radius width is 5.33~km, about 4.1 times the product-limit width. The median mass changes by only $0.04\,M_\odot$. The uncertainty changes substantially while the central mass remains stable. The adopted interval is narrower than the fully dependent endpoint. The green and orange prescriptions change both the correlation treatment and the common displacement, so this ordering does not isolate either effect.

The separate density-agreement diagnostic in Section~\ref{sec:tempering} gives $\rho=0.23^{+0.38}_{-0.17}$. This broad distribution describes agreement among the supplied densities. The J0030 estimate retains the prescribed uniform correlation weights rather than using this diagnostic to infer independence from posterior overlap.

The combined posterior samples, together with all scripts and configuration files needed to reproduce the analysis, are publicly available via our Zenodo repository at \url{https://doi.org/10.5281/zenodo.19222905}.

\subsection{Statistical consistency ranking}
\label{sec:consistencyranking}

An important consequence of this Bayesian formulation is that we can compute, for each measurement $i$, the posterior probability $P_i$ that measurement $i$ is ``good'', or in other words statistically consistent with the ensemble under the mixture model. In the present formulation, this probability is computed by marginalizing over the nuisance parameters of the mixture. At fixed $\rho$ we write
\begin{align}
P_i(\rho)
&=
\int d^2x
\int_0^1 dp
\int d\boldsymbol{\sigma}\,
P_\rho(\boldsymbol{x},p,\boldsymbol{\sigma}\mid D)
\nonumber\\
&\quad \times
\frac{
p\,q_i(\boldsymbol{x})
}{
p\,q_i(\boldsymbol{x})
+
(1-p)\,\bar B_i(\boldsymbol{x}\mid \boldsymbol{\sigma})
},
\label{eq:consistency}
\end{align}
and report $P_i=\int_0^1 d\rho\,\pi_\rho(\rho)P_i(\rho)$. Here $P_\rho(\boldsymbol{x},p,\boldsymbol{\sigma}\mid D)\propto\pi_x\pi_\eta L_\rho(\boldsymbol{x}\mid\boldsymbol{\eta})$ is the joint posterior of the tempered pool (Eq.~\ref{eq:tempered_pool}), $q_i$ is the KDE of input $i$, and $\bar B_i$ is its bad component (Eq.~\ref{eq:goodandbad}), with $\alpha_i$ and $\boldsymbol{\Delta}_i$ already marginalized. The common systematic $\boldsymbol{\Sigma}_{\rm sys}$ is applied after the pool is formed and does not enter $P_i$. Thus, $P_i$ quantifies how strongly the combined posterior favors interpreting measurement $i$ as belonging to the ``good'' component rather than the broadened-and-shifted bad component.

Table~\ref{tab:probabilitygood} summarizes the posterior probabilities $P_i$ for each of the eight measurements we analyzed in this paper. The ordering and values of $P_i$ are roughly consistent with the overlap of our $M$--$R$ posterior distribution with the original measurement $i$, as seen in Figure~\ref{fig:3x3PosteriorvsIndividual}. The highest $P_i$ is from the Kini26 PDT--U model ($P_i = 0.5302$), which is the most consistent with our collective posterior combination, while the lowest $P_i$ is given by the Vinciguerra24 ST+PST model ($P_i = 0.1873$), the least consistent with our combined posterior.

The Vinciguerra24 ST+PST posterior lies at higher mass and radius ($M\approx1.9\,M_\odot$, $R\approx15$~km; Table~\ref{tab:summary_table}) than most of the ensemble. As discussed in \citet{Vinciguerra24}, such behavior may arise when the radius posterior approaches the upper limit of the prior range, which in turn pushes the inferred mass upward through the compactness constraint. We do not re-evaluate that explanation here. In our framework, this offset gives ST+PST the smallest overlap with the combined posterior and therefore the lowest $P_i$.

\begin{table}
\centering
\small
\setlength{\tabcolsep}{4pt}
\begin{tabular}{@{}lc@{}}
\hline\hline
Model & $P_i$ \\
\hline
Kini26: PDT--U          & 0.5302 \\
Miller19: 3 spots       & 0.4743 \\
Vinciguerra24: PDT--U   & 0.4557 \\
Miller19: 2 spots       & 0.4441 \\
Riley19: ST+PST         & 0.4190 \\
Vinciguerra24: ST--U    & 0.3550 \\
Vinciguerra24: ST+PDT   & 0.3482 \\
Vinciguerra24: ST+PST   & 0.1873 \\
\hline\hline
\end{tabular}
\caption{
Posterior probability $P_i$ that each input analysis belongs to the ``good'' component of the mixture, evaluated with the adopted uniform correlation prior. The greater the agreement between an individual $M$--$R$ posterior and the combined posterior, the closer $P_i$ is to 1.
}
\label{tab:probabilitygood}
\end{table}

\section{Discussion and Conclusion}
\label{sec:conclusion}

In this work, we developed and applied a hotspot-model-agnostic Bayesian combination framework to account for the dominant model-choice systematic in NICER pulse-profile inferences for PSR~J0030+0451. Following a Bayesian framework for handling systematic uncertainties \citep{press1996,Lahav2000, Hobson2002, Bernal_2018, phillips2025}, we treated each published hotspot-dependent inference as an imperfect constraint that may or may not be statistically consistent with the underlying stellar parameters, and we marginalized over the probability that a given input posterior is ``good'' rather than assuming all measurements are intrinsically correct. To faithfully represent the strongly non-Gaussian structure of published posteriors, we used kernel density estimates (KDEs) in place of Gaussian approximations, with bandwidths chosen using Silverman’s multivariate suggestion. Although the results are presented in $M$--$R$, the combination is performed in $M$--$C$, such that the bad components and the common systematic, which are diagonal in $M$ and $C$, extend along lines of constant compactness.

We combined eight NICER-based $M$--$R$ posteriors for J0030+0451 (including analyses that also incorporate XMM-Newton constraints), spanning a broad set of physically motivated hotspot morphologies. These analyses share data from NICER and XMM-Newton observations, so we tempered their combined contribution to allow for unknown dependence (Section~\ref{sec:tempering}) and marginalized over a common systematic displacement (Section~\ref{sec:groupsys}). The use of a single parameter for correlation tempering simplifies the relationship between these analyses. Because the degree of shared information cannot be measured from the published posteriors alone, $\rho$ is assigned a prior rather than inferred from the data; our uniform prior spans the full range from completely uncorrelated ($\rho = 0$) to completely correlated ($\rho = 1$), which is reflected in the width of our combined posterior.

The resulting joint posterior provides a single summary of the J0030 mass and radius for use when no independent discriminator exists among competing hotspot configurations.
Our combined posterior distribution is centered at
\[
M_0 = 1.53^{+0.31}_{-0.27}\,M_\odot,\;
R_0 = 13.44^{+2.95}_{-2.38}\,\mathrm{km},\;
C_0 = 0.169^{+0.018}_{-0.021}.
\]
The central 95\% marginal intervals are $[1.03,2.10]\,M_\odot$ and $[8.79,21.47]$~km. The pool is broader than most individual inputs: it retains the spread between the lower- and higher-mass solutions rather than selecting one of them. The tails of this posterior run up against the boundaries of the grid over which it is computed, making the result sensitive to the range of the grid. (Section~\ref{sec:groupsys}) Treating the inputs as independent ($\rho=0$) would instead give a radius interval about 4.1 times narrower (Section~\ref{sec:temper_results}).

A further outcome of the Bayesian formulation is an internally defined consistency score, $P_i$, for each input posterior, which quantifies agreement of the input posterior with the combined constraint. Under this metric, the PDT--U posterior from \citet{Kini2026} has the highest $P_i$, whereas the XMM-background-constrained ST+PST posterior in \citet{Vinciguerra24} receives the lowest $P_i$, indicating the greatest tension with the combined posterior (Sec.~\ref{sec:consistencyranking}; Table~\ref{tab:probabilitygood}).

We emphasize that the per-measurement consistency score $P_i$ is an internal diagnostic within the adopted Bayesian combination method, rather than a metric of physical correctness for each model; it quantifies agreement with the ensemble after marginalizing over unknown systematics. It is not the probability that a given hotspot morphology is physically correct, nor should it be used to argue for astrophysical plausibility. This analysis also assumes equal weight between the eight input posteriors in the combination. For downstream applications (e.g., EoS inference), the intended use of this work is to adopt the combined posterior samples as a summary of the current NICER-informed constraints on J0030+0451 when no external discriminator among hotspot configurations is available.

Regarding the implications for the dense-matter EoS, we defer a full EoS Bayesian inference to future work. Here we instead provide an approximate, EoS-agnostic assessment of what the combined PSR~J0030+0451 constraint can imply, following the strategy developed by \citet{Huang:2024wig,Huang:2025mrd}. Our focus is on quantities of broad utility for EoS studies, in particular the canonical-radius $R_{1.4}$ and tidal deformability $\Lambda_{1.4}$ of a $1.4\,M_\odot$ neutron star, which can be estimated without performing a computationally intensive Bayesian EoS analysis.

\begin{figure*}
    \centering
    \begin{minipage}{0.47\linewidth}
        \centering
        \includegraphics[width=1\linewidth]{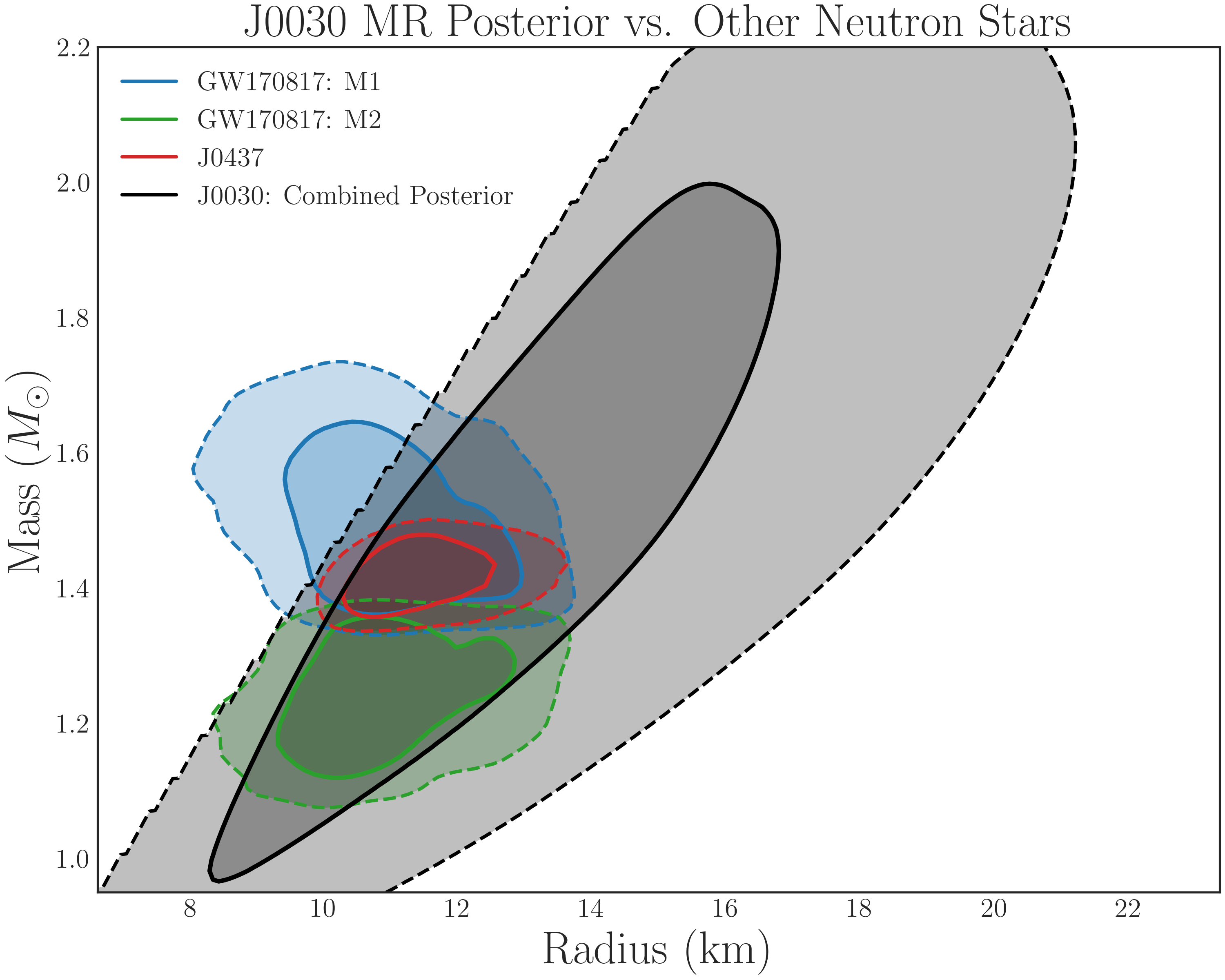}
        \label{fig:overotherMR}
    \end{minipage}%
    \begin{minipage}{0.5\linewidth}
        \centering
        \includegraphics[width=1\linewidth]{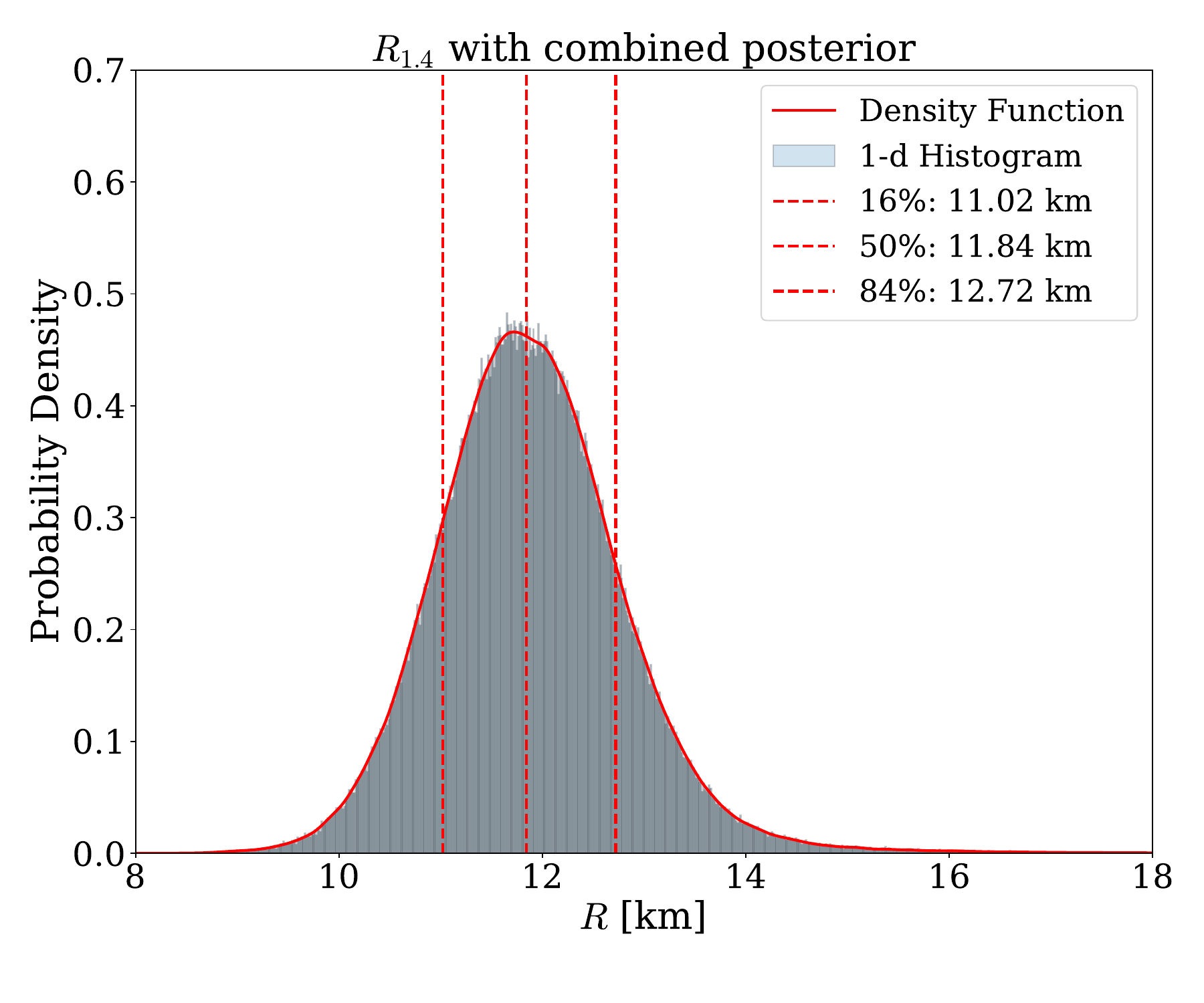}
        \label{fig:R14}
    \end{minipage}%
    \caption{
Left: Mass--radius posteriors for GW170817 from the EoS-insensitive analysis, overlaid with the posterior for PSR~J0437--4715 and our combined posterior for PSR~J0030+0451.
Dark (light) shading indicates the 68\% (95\%) credible region.
Right: Posterior distribution of the canonical radius $R_{1.4}$ obtained by combining our PSR~J0030+0451 constraint with PSR~J0437--4715; dashed vertical lines mark the 16\%, 50\%, and 84\% quantiles.
The density curve is estimated with KDE.
}
    \label{fig:EOS_ind}
\end{figure*}

A key practical point is that the combined PSR~J0030+0451 posterior reported here has support near $M \simeq 1.4\,M_\odot$ within its $1\sigma$ credible interval. This places J0030+0451 in the same mass regime as several other high-impact constraints, most notably PSR~J0437--4715 and the binary neutron star merger GW170817 (see left panel of Fig.~\ref{fig:EOS_ind}). Relative to these systems, our combined $M$--$R$ posterior for PSR~J0030+0451 is centered at a slightly higher mass and extends over a large range of radii. These discrepancies are particularly consequential when translating the posterior into an inferred canonical radius $R_{1.4}$ for a $1.4\,M_\odot$ neutron star.

Our inferred $R_{1.4}$, obtained by combining the information from PSR~J0437--4715 using the Scenario~1 method of \citet{Huang:2024wig}, is shown in right panel of Fig.~\ref{fig:EOS_ind}. In this framework, we find
\begin{equation}
    R_{1.4} = 11.84^{+0.88}_{-0.83}\ \mathrm{km},
\end{equation}
where the uncertainties denote the central 68\% credible interval. The central 95\% interval is $[10.22,13.80]$~km, with a width of 3.58~km, compared with 2.92--3.42~km for the Scenario~1 results in Table~2 of \citet{Huang:2024wig}. At matched credibility, our interval is therefore slightly broader. The interpolation is weighted toward PSR~J0437--4715, whose mass is close to $1.4\,M_\odot$; its median interpolation weight is approximately 0.88. The width of this conditional $R_{1.4}$ distribution is consequently not determined by the J0030 radius marginal alone.

When we further combine PSR J0030+0451 and PSR~J0437--4715 with GW170817, the resulting multimessenger constraint on the tidal deformability is
\begin{equation}
    \Lambda_{1.4} = 271^{+285}_{-108},
\end{equation}
using the method and empirical radius--tidal-deformability relation of \citet{Huang:2025mrd}. Here the central value is obtained by transforming the mode of the joint radius distribution, while the bounds are the 5th and 95th percentiles. This posterior has a slightly higher central value and a broader 90\% credible interval compared to the headline result in \citet{Huang:2025mrd}. These constraints are broadly compatible with the GW170817 measured inference and do not require a phase transition or twin-star interpretation.

\section*{Software and Third Party Data Repository Citations}
\texttt{SciPy}, \citet{Scipy}; \texttt{seaborn}, \citet{Waskom2021}; \texttt{UltraNest}, \citet{Ultranest}; \texttt{CompactObject}, \citet{compactObject}. Posterior samples used in this work: \citet{riley19c,J0030zenodoBRAVO,miller19j0030,kininicer,choudhury24j0437,Abbott_2018_EOS_GW}, the reproduction scripts, posterior samples, and headline result posteriors for this work are publicly available on Zenodo at \url{https://doi.org/10.5281/zenodo.19222905}. 

\begin{acknowledgments}
The authors are grateful to Michael Nowak for directing us to key references in the original literature. C.H.\ also thanks Cole Miller for insightful discussions and comments. A.C.\ and C.H.\ acknowledge support from NSF grant AST-2308111. A.C.\ additionally acknowledges support from NSF grant DMS-2235457 and NASA grant 80NSSC24K1095. R.O.C.\ acknowledges support from the M.R.\ Metzger Family Foundation's Undergraduate Research Fellowship. 

The authors used generative artificial-intelligence
tools, including OpenAI Codex and Anthropic Claude,
to assist with numerical cross-checks, code development,
and language editing during the analysis and preparation of this manuscript. All scientific results, interpretations, and final manuscript content were independently
reviewed and verified by the authors.

\end{acknowledgments}
\bibliography{ref}{}
\bibliographystyle{aasjournal}

\end{document}